\documentclass[10pt,conference,review,anonymous]{IEEEtran}

\usepackage[pdftex]{graphicx}
\usepackage[usenames,dvipsnames,table,xcdraw]{xcolor}
\usepackage{array}
\usepackage{multirow}
\usepackage[normalem]{ulem}
\useunder{\uline}{\ul}{}
\usepackage{booktabs}
\usepackage{supertabular,booktabs}
\usepackage{longtable}
\usepackage{lscape}
\usepackage{cite}
\usepackage{url}
\usepackage{hyperref}
\usepackage{dirtytalk}
\usepackage{comment}
\usepackage{makecell}
\usepackage{svg}
\usepackage{amsmath}
\usepackage{float}
\usepackage[caption=false]{subfig}
\usepackage{tabularx}
\usepackage{balance}
\usepackage{nameref}
\usepackage[most]{tcolorbox}
\usepackage{multirow}
\usepackage{booktabs}

\graphicspath{{Images/}}

\definecolor{lightblue}{RGB}{0,0,100}
\definecolor{purplish}{HTML}{D8DFE3}
\definecolor{purplishlight}{HTML}{EBEFF3}
\definecolor{purplishdark}{HTML}{FF7F50}

\newtcolorbox{MyBox}{
  colback=white,
  colframe=lightblue,
  fonttitle=\bfseries,
  coltitle=black,
  sharp corners,
  boxrule=1pt,
  left=5pt,
  right=5pt,
  top=5pt,
  bottom=5pt,
  breakable
}

\newtcolorbox[auto counter,number within=section]{rqbox}[2]{
    nameref=#1,
    title=\small{#1},
    enhanced,
    attach boxed title to top left={yshift=-6pt, xshift=8pt},
    boxed title style={size=small,boxsep=1pt},
    colframe=purplishdark,
    colback=white,
    colbacktitle=purplishdark,
    boxsep=2pt,
    left=2pt,
    right=2pt,
    top=6pt,
    bottom=2pt,
    middle=2pt
}

\begin{document}

\title{Collaborative and AI-Supported Requirements Elicitation: An Empirical Study}

\author{

\IEEEauthorblockN{Manoel Salgado Neto}
\IEEEauthorblockA{CESAR School\\
Recife, PE, Brazil \\
mvsn@cesar.school} 

\and

\IEEEauthorblockN{Allan Araujo}
\IEEEauthorblockA{CESAR School\\
Recife, PE, Brazil \\
arsa@cesar.school}

\and

\IEEEauthorblockN{Ronnie de Souza Santos}
\IEEEauthorblockA{University of Calgary\\
Calgary, AB, Canada \\
ronnie.desouzasantos@ucalgary.ca}

}

\maketitle

\IEEEpeerreviewmaketitle

\begin{abstract}
Requirements elicitation requires stakeholders to communicate needs, negotiate priorities, and collaboratively construct knowledge that can be transformed into software requirements artifacts. Recent advances in LLMs have created opportunities to support these activities through AI-assisted collaboration and automated artifact generation. However, limited empirical evidence is available regarding how AI-supported collaborative environments influence requirements elicitation outcomes. In this study, we conducted a mixed-method controlled experiment comparing four requirements elicitation approaches: collaborative elicitation without AI support, collaborative elicitation supported by the Strateegia platform and its GPT-powered Writer applet, direct requirements generation using a Large Language Model, and requirements generation from collaborative discussion transcripts using a Large Language Model. We evaluated the resulting artifacts using quality criteria derived from ISO/IEC/IEEE 29148 and collected participant perceptions regarding the elicitation process. Our findings indicate that approaches combining stakeholder collaboration and AI-supported synthesis produced the highest-rated requirements artifacts and were perceived as clearer and easier to execute than traditional collaborative elicitation. The results suggest that generative AI can support the transformation of collaboratively generated knowledge into structured requirements documentation while preserving the value of stakeholder participation. We discuss implications for AI-supported requirements elicitation and human-AI collaboration in Requirements Engineering.
\end{abstract}


\section{Introduction}
\label{sec:introduction}

Requirements elicitation is a common activity in software development, involving the identification, discussion, and documentation of stakeholder needs that guide subsequent design and implementation activities~\cite{sommerville2011software,pressman2005software}. This process typically involves multiple stakeholders, including clients, users, developers, and managers, who may possess different goals, expertise, and perspectives regarding the system under development~\cite{sommerville2011software,nuseibeh2000requirements}. To support coordination among these stakeholders, collaborative platforms have been proposed as mechanisms to facilitate collective knowledge construction and stakeholder interaction. More recently, Large Language Models (LLMs) have been incorporated into a variety of software engineering activities due to their ability to interpret, summarize, and generate natural language content~\cite{brown2020language,vaswani2017attention}. These capabilities create opportunities to support requirements elicitation and documentation activities in collaborative environments.

Despite the availability of such tools, requirements elicitation remains associated with coordination challenges among stakeholders~\cite{davey2015requirements}. Stakeholders may express needs using ambiguous, incomplete, or evolving language, while developers are responsible for translating these discussions into formal requirements specifications. Differences in interpretation, changing expectations, incomplete participation, and misaligned understanding may contribute to requirements documents containing ambiguities, inconsistencies, or omissions~\cite{davey2015requirements,pacheco2018requirements}. Such issues have been associated with rework, increased development effort, project delays, and software systems that do not fully address stakeholder needs~\cite{sommerville2011software,pressman2005software,iqbal2020requirements}.

Regarding potential solutions, collaborative platforms and LLM-based systems provide mechanisms that may offer relevant support for requirements elicitation activities. Collaborative environments facilitate stakeholder participation and the externalization of knowledge, while LLMs can assist in organizing, synthesizing, and transforming information generated during discussions~\cite{arora2024advancing,abbasi2025towards}. Recent work has investigated the use of generative AI in software engineering activities, including requirements engineering~\cite{arora2024advancing,abbasi2025towards}. However, limited empirical evidence is available regarding how AI-supported collaborative environments influence requirements elicitation outcomes. In particular, little is known about how collaboratively generated knowledge can be transformed into requirements artifacts through generative AI, or how such approaches compare with traditional collaborative elicitation and LLM-based artifact generation approaches in terms of artifact quality and participant experience~\cite{arora2024advancing,abbasi2025towards}.

To address this gap, we investigate the following research question:

\begin{quote}
\textbf{\textit{How does an AI-supported collaborative platform influence software requirements elicitation and the production of requirements artifacts compared with alternative elicitation approaches?}}
\end{quote}

To answer this question, we conducted a mixed-method controlled experiment comparing four approaches to requirements elicitation and artifact generation: collaborative elicitation without AI support, collaborative elicitation supported by the \textit{Strateegia} \footnote{https://strateegia.digital/en} platform and its GPT-powered \textit{Writer} applet, direct requirements generation using a Large Language Model, and requirements generation from collaborative discussion transcripts using a Large Language Model. The study combined artifact quality assessment, participant perceptions, and qualitative observations to investigate both process and outcome differences among the approaches.

This paper makes four contributions: (1) an empirical comparison of collaborative, AI-supported, and LLM-based requirements elicitation approaches, (2) an evaluation of generative AI mechanisms for transforming stakeholder discussions into requirements artifacts, (3) evidence regarding participant experiences with AI-supported collaborative elicitation, and (4) implications for the adoption of collaborative platforms and generative AI in Requirements Engineering.

\section{Background}
\label{sec:background}

This section provides the background necessary to understand the context of this study. We begin by discussing requirements elicitation, collaborative platforms, and the use of Artificial Intelligence (AI) to support requirements engineering activities, followed by a presentation of existing tools and approaches that integrate AI into requirements-related tasks.

\subsection{Requirements Elicitation, Collaborative Platforms, and Artificial Intelligence}

Requirements elicitation is a central activity in Requirements Engineering (RE), concerned with identifying, understanding, and documenting stakeholder needs, expectations, and constraints that will guide software development efforts \cite{sommerville2011software,nuseibeh2000requirements}. Traditional elicitation approaches commonly rely on interviews, questionnaires, workshops, observation, and other stakeholder-centered techniques \cite{pacheco2018requirements}. Over time, RE research and practice have increasingly emphasized collaborative approaches that promote interaction among multiple stakeholders and support the collective construction of knowledge. Methods such as Joint Application Development (JAD) and the Nominal Group Technique (NGT) encourage collective discussion, negotiation, and decision-making during elicitation activities. More recently, technology-assisted approaches have emerged, incorporating AI, Natural Language Processing (NLP), and requirements mining techniques to support parts of the elicitation process \cite{marques2024using,arora2024advancing}. These developments reflect continued efforts to improve stakeholder participation, reduce elicitation effort, and support the production of higher-quality requirements artifacts.

Collaborative digital platforms provide socio-technical environments that support collective knowledge construction and co-authoring activities. These platforms enable both synchronous and asynchronous participation, allowing stakeholders to discuss ideas, exchange perspectives, and collaboratively construct shared artifacts \cite{neves2022strateegia}. As an example of such environments, the collaborative platform \textit{strateegia} provides structured support for stakeholder interaction through guided discussion and decision-making journeys. The platform is designed to facilitate collective reflection and knowledge construction among participants engaged in complex problem-solving activities.

Recent advances in AI have expanded the capabilities available within collaborative environments. AI has evolved from symbolic and statistical approaches toward connectionist and generative models enabled by advances in deep learning and Transformer-based architectures \cite{lecun2015deep,vaswani2017attention}. Within this context, NLP has enabled the development of Large Language Models (LLMs) capable of interpreting, generating, summarizing, and transforming textual information in context \cite{brown2020language}. Recent studies suggest that these models can support requirements-related activities such as requirements elicitation, information extraction, requirements interpretation, requirements formalization, quality analysis, and artifact generation \cite{marques2024using,arora2024advancing,abbasi2025towards}.

The integration of LLMs into requirements engineering has motivated the development of human-AI collaboration approaches in which AI assists analysts and stakeholders throughout elicitation, analysis, specification, and validation activities \cite{abbasi2025towards,arora2024advancing}. Following this trend, \textit{strateegia} incorporates the \textit{Writer} applet, which uses GPT-4.1 and GPT-4.1 Mini models to support text generation and knowledge synthesis from participant contributions. Such capabilities illustrate how collaborative environments can integrate generative AI mechanisms to assist participants during knowledge-intensive activities.

Despite growing interest in AI-supported requirements engineering, empirical evidence regarding the use of LLMs in collaborative requirements elicitation settings remains limited \cite{abbasi2025towards,arora2024advancing}. Existing studies have primarily focused on evaluating AI support for individual requirements engineering tasks, such as requirements generation, specification, analysis, and validation. Comparatively less attention has been given to understanding how collaboratively generated knowledge can be transformed into requirements artifacts through generative AI, or how AI-supported collaborative elicitation approaches compare with more traditional elicitation strategies.

The literature therefore points to a convergence between Requirements Engineering, collaborative digital platforms, and Artificial Intelligence. While each of these areas has matured independently, limited empirical evidence is available regarding how collaborative knowledge construction and generative AI can be combined to support requirements elicitation activities. This study contributes to this discussion by investigating the use of the collaborative platform \textit{strateegia} and its GPT-powered \textit{Writer} applet as complementary mechanisms to support collaborative requirements elicitation.

\subsection{AI-Supported Requirements Engineering Tools}
\label{sec:tools}

Several tools have been proposed to support Requirements Engineering activities throughout the software development lifecycle. Existing solutions differ in their emphasis on requirements management, stakeholder collaboration, traceability, and AI support.

\begin{itemize}

\item \textbf{IBM Engineering Requirements Management DOORS Next.} DOORS Next is a mature platform widely adopted in large-scale and highly regulated domains. Its primary capabilities include requirements traceability, version control, and change management. The platform focuses primarily on requirements management after elicitation and provides limited support for collaborative stakeholder discussions.

\item \textbf{Jama Connect.} Jama Connect combines requirements management, collaboration, traceability, and workflow support. While it facilitates stakeholder participation and standards compliance, its emphasis remains on managing and maintaining requirements artifacts rather than supporting collaborative elicitation activities.

\item \textbf{aqua Cloud.} aqua Cloud incorporates AI capabilities to support requirements writing, artifact review, and test case generation. AI support is primarily focused on assisting artifact creation and refinement, functioning largely as a writing assistant rather than supporting structured collaborative elicitation.

\item \textbf{ReqView and Valispace.} ReqView and Valispace are lightweight solutions that provide traceability and integration with engineering workflows. However, they offer limited support for structured stakeholder discussions and AI-assisted synthesis of elicitation outcomes.

\end{itemize}

These tools illustrate the range of support currently available for Requirements Engineering activities. Existing solutions provide capabilities related to requirements management, collaboration, traceability, standards compliance, and AI-assisted writing. However, support for combining structured stakeholder discussions with generative AI to synthesize collaborative interactions into requirements artifacts remains comparatively limited \cite{arora2024advancing,abbasi2025towards}. To investigate this context, we use the collaborative platform \textit{strateegia}. Unlike the tools discussed above, \textit{strateegia} was designed to support structured collaborative processes through discussion and decision-making journeys involving multiple participants. The platform also includes AI-enabled applets capable of generating textual artifacts from participant contributions. Although \textit{strateegia} was not developed specifically for Requirements Engineering, these characteristics make it a suitable environment for investigating how collaborative platforms and generative AI can support requirements elicitation activities.

\section{Methodology}
\label{sec:method}

\begin{figure*}[t]
\centering
\includegraphics[width=0.9\textwidth]{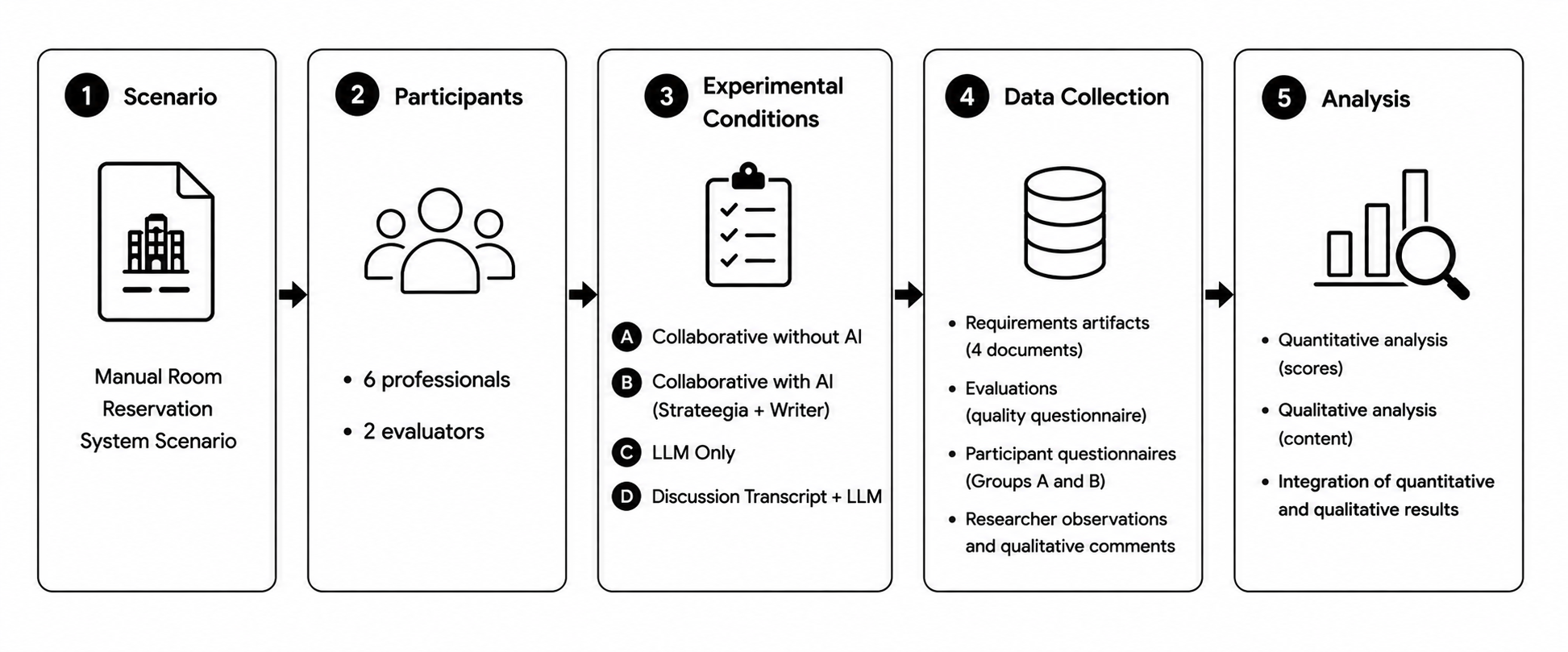}
\caption{Overview of the study methodology.}
\label{fig:methodology_overview}
\end{figure*}

We conducted a quasi-experiment \cite{kampenes2009systematic, wohlin2012experimentation} to investigate the use of a collaborative platform supported by generative AI for software requirements elicitation. Figure~\ref{fig:methodology_overview} provides an overview of the study design. We compared different approaches for eliciting and documenting software requirements from the same problem scenario, combining quantitative measures related to artifact quality and participant perceptions with qualitative evidence collected through post-activity reflections.

\subsection{Context and Platform}

We designed a controlled requirements elicitation exercise in which participants were asked to produce a requirements artifact from the same fictitious problem scenario. \\

\noindent \textbf{Fictitious problem scenario.} We created a scenario describing an organization that manages meeting rooms, auditoriums, and laboratories through a manual reservation process based on emails and spreadsheets. The scenario included scheduling conflicts, overlapping reservations, limited traceability, and the absence of standardized approval procedures. Participants were asked to elicit requirements for a hypothetical software system intended to address these challenges. Rather than developing a working system, participants were expected to produce a requirements artifact based exclusively on the information provided in the scenario. We instructed participants to organize the elicited requirements according to the structure recommended by ISO/IEC/IEEE 29148, including system scope, functional requirements, non-functional requirements, constraints, risks, prioritization decisions, and a minimum viable product (MVP) definition. \\

\begin{figure*}[t]
\centering
\includegraphics[width=0.9\textwidth]{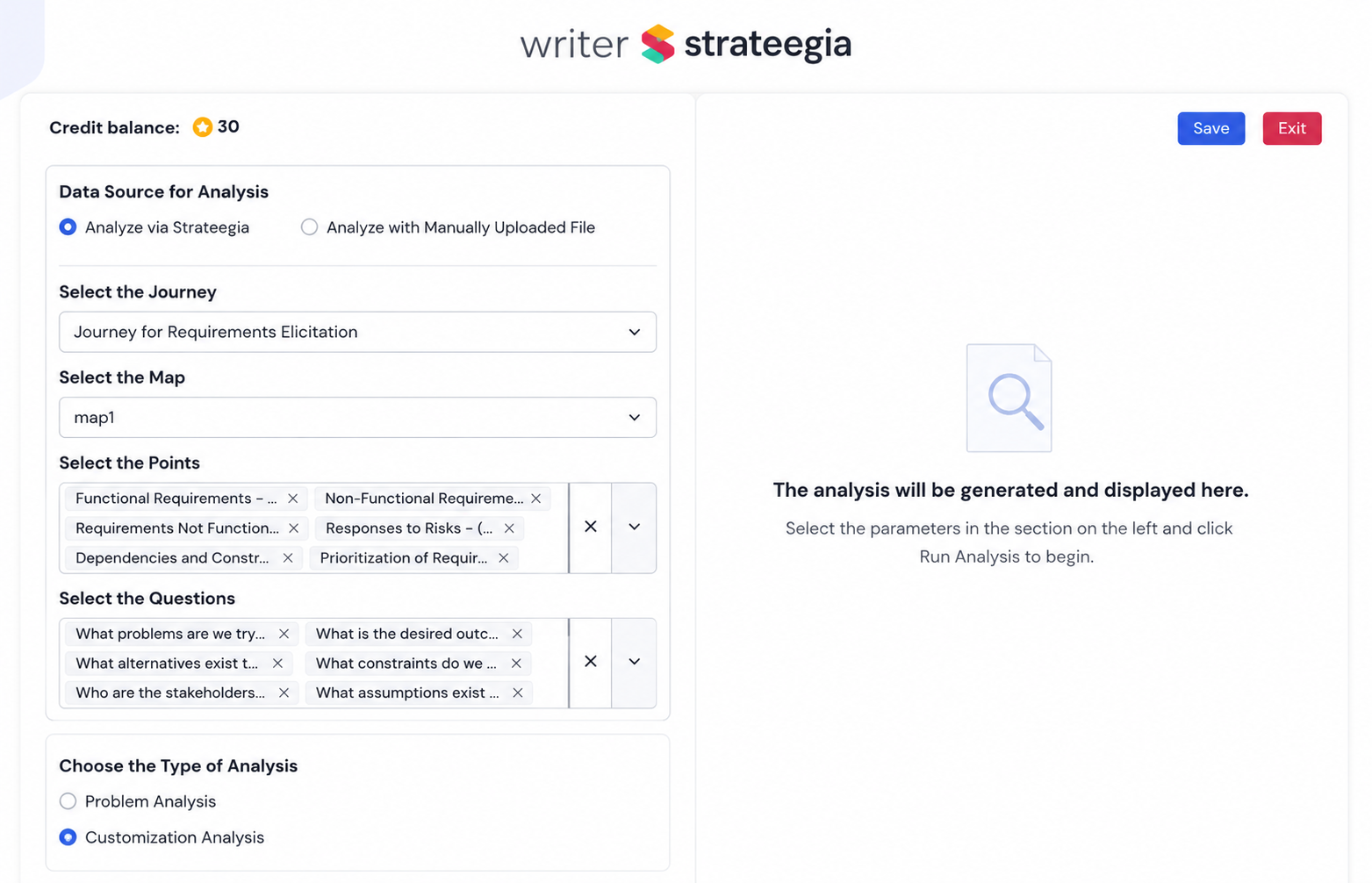}
\caption{Writer Applet Interface in the Strateegia platform.}
\label{fig:writer}
\end{figure*}

\noindent \textbf{Strateegia platform.} For the AI-supported collaborative condition, we used the \textit{Strateegia} platform~\cite{neves2022strateegia}. We selected Strateegia because it provides a structured environment designed to support collaborative knowledge construction, creative discussion, and the systematic recording of participant contributions. Unlike generic communication tools such as videoconferencing platforms or instant messaging applications, Strateegia organizes participant interactions through collaboration journeys composed of predefined interaction points. These interaction points can support activities such as debate, evaluation, decision-making, monitoring, meetings, and announcements, allowing participants to collectively explore ideas, discuss alternatives, and converge toward shared outcomes. The platform was designed to support both synchronous and asynchronous participation, enabling contributors to interact throughout different stages of a collaborative process. Figure~\ref{fig:writer} presents the Writer applet interface used during the study. The motivation for selecting Strateegia was the availability of the \textit{Writer} applet, a GPT-powered text generation component based on GPT-4.1 and GPT-4.1 Mini. Writer was originally designed to synthesize participant contributions and generate coherent textual artifacts from collaborative discussions. In this study, we extended its use to the context of Requirements Engineering by investigating its ability to transform collaboratively generated knowledge into structured requirements documentation. \\

\noindent \textbf{ISO/IEC/IEEE 29148 requirements structure.} To guide the organization of the generated requirements artifacts, we adopted the recommendations provided by ISO/IEC/IEEE 29148 \cite{iso2018iec}, the international standard that defines concepts, processes, and documentation practices for Requirements Engineering. The standard provides guidance regarding the specification of stakeholder and system requirements and recommends documenting information such as stakeholder needs, system context, functional requirements, quality requirements, assumptions, constraints, risks, rationale, and prioritization information. Rather than prescribing a single document template, the standard identifies key categories of information that contribute to complete, traceable, and verifiable requirements specifications. To align the elicitation process with these recommendations, we designed a requirements elicitation journey containing stages for problem understanding, stakeholder identification, scenario exploration, functional requirements, non-functional requirements, risks, constraints, clarification activities, and prioritization decisions. After participants completed the collaborative activities, the Writer applet processed the collected contributions and generated a requirements artifact organized according to these categories. The resulting document synthesized the collaboratively generated knowledge into a structured requirements specification without further manual modification.

\subsection{Experimental Design}

We adopted a between-subjects experimental design in which participants were exposed to a single experimental condition. We selected this design to reduce learning and familiarity effects that could influence the results if participants completed the same elicitation task multiple times. Since all conditions used the same problem scenario, exposure to one condition could influence performance in subsequent conditions and compromise the comparability of the results~\cite{wohlin2012experimentation}. We designed the experiment to compare different combinations of stakeholder collaboration and AI support during requirements elicitation. Four experimental conditions were evaluated:

\begin{itemize}

\item \textbf{Condition A (Collaborative without AI).} Participants collaboratively elicited requirements using activities inspired by Joint Application Development (JAD) and the Nominal Group Technique (NGT). Discussions, prioritization activities, and document production were performed without AI support.

\item \textbf{Condition B (Collaborative with AI).} Participants completed the elicitation process using the Strateegia platform. Collaborative discussions and prioritization activities were conducted through the platform, and the final requirements artifact was generated using the Writer applet.

\item \textbf{Condition C (LLM Only).} The main author used a Large Language Model to generate a requirements artifact directly from the problem scenario without collaborative activities. A second author followed the process closely and conflicts were solved through discussion and consensus meetings.

\item \textbf{Condition D (Discussion Transcript + LLM).} We provided the discussion transcript produced during Condition A to a Large Language Model, which generated a requirements artifact based on the information contained in the collaborative discussion.

\end{itemize}

Conditions C and D were included as baseline AI-assisted artifact generation approaches rather than collaborative participant groups. All conditions received the same problem scenario and were expected to produce requirements artifacts following the same ISO/IEC/IEEE 29148 structure.

\subsection{Participants}

We recruited participants through convenience and snowball sampling \cite{baltes2022sampling}. The study involved six technology professionals distributed across two collaborative elicitation groups and two independent requirements engineering specialists responsible for artifact evaluation. Group A and Group B each included three software professionals representing complementary project perspectives (Product Owner, Quality Assurance, and Software Developer). Evaluator profiles and selection criteria are summarized in Table~\ref{tab:participants}. Given the exploratory nature of the experiment, the objective was not statistical generalization but rather the comparative evaluation of different requirements elicitation approaches under controlled conditions.

\begin{table}[!ht]
\caption{Participant profiles}
\label{tab:participants}
\centering
\footnotesize
\begin{tabular}{p{1.5cm} p{1.8cm} p{2.0cm} p{2.0cm}}
\toprule
\textbf{Group} & \textbf{Role} & \textbf{Profile} & \textbf{Inclusion} \\
\midrule

\multirow{3}{*}{A}
& Product Owner & Software projects & IT experience \\
& QA & Software projects & IT experience \\
& Developer & Software projects & IT experience \\
\midrule

\multirow{3}{*}{B}
& Product Owner & Software projects & IT experience \\
& QA & Software projects & IT experience \\
& Developer & Software projects & IT experience \\
\midrule

Evaluators
& RE Specialist (2)
& Industry RE experience
& Demonstrated RE experience \\
\bottomrule
\end{tabular}
\end{table}

\subsection{Data Collection}

We collected data related to both the requirements artifacts produced during the experiment and participants' perceptions of the elicitation process. Each experimental condition produced a single consolidated requirements artifact organized according to the structure recommended by ISO/IEC/IEEE 29148. This process resulted in four requirements documents (A, B, C, and D), which constituted the primary outputs of the experiment and the basis for the artifact quality evaluation.

The recruited evaluators assessed all generated documents using a questionnaire derived from quality attributes associated with ISO/IEC/IEEE 29148. The evaluation instrument included measures of completeness, clarity, consistency, absence of ambiguity, level of detail, traceability, and overall quality. This process produced evaluator ratings for each quality attribute and each generated artifact.

To assess process quality, we administered a post-activity questionnaire to participants in the collaborative conditions (Groups A and B). The questionnaire captured perceptions regarding process clarity, ease of execution, collaboration flow, perceived cognitive effort, and the adequacy of the supporting tools. This process produced quantitative ratings and qualitative comments regarding participant experiences with the elicitation process.

In addition to the questionnaires, we collected researcher observations recorded throughout the experiment. Together with the open-ended responses, these observations constituted the qualitative dataset used to support the interpretation of the quantitative findings and provide additional insight into participant experiences and challenges encountered during the elicitation activities.

\subsection{Data Analysis}

We adopted a mixed-method analysis strategy combining quantitative and qualitative evidence. Quantitative analysis focused on the scores assigned by the evaluators to the generated requirements artifacts and on the responses provided by participants in the process evaluation questionnaire.

For artifact quality, we calculated the mean evaluator score for each quality attribute, namely completeness, clarity, consistency, absence of ambiguity, level of detail, traceability, and overall quality. These scores were computed for each generated document (A, B, C, and D) and used to compare the outcomes of the four elicitation approaches.

For process quality, we calculated the mean participant score for each questionnaire item, including process clarity, ease of execution, collaboration flow, perceived cognitive effort, and tool adequacy. Given the exploratory nature of the study and the limited number of participants and evaluators, the quantitative analysis was restricted to descriptive statistics and should be interpreted as providing descriptive indicators rather than statistically generalizable findings.

Qualitative analysis focused on the open-ended questionnaire responses and researcher observations collected throughout the experiment. Using content analysis \cite{hsieh2005three}, we reviewed these materials iteratively to identify recurring perceptions regarding strengths, limitations, collaboration dynamics, challenges, and the role of AI during the elicitation process. Similar observations were grouped into higher-level categories that were subsequently used to support the interpretation of the quantitative findings.

Finally, we integrated the quantitative and qualitative results during the interpretation stage. Quantitative results were used to identify differences among elicitation approaches, while qualitative evidence was used to provide context and possible explanations for the observed patterns.

\subsection{Threats to Validity}

The findings of this study should be interpreted in light of the study design and methodological choices adopted. First, the experiment involved a small number of participants and was conducted within a limited time window. Although appropriate for an exploratory investigation, different participant populations, team compositions, or longer elicitation sessions may produce different results. Second, the study relied on a single elicitation scenario involving a room reservation management context. Different domains, problem complexities, or stakeholder contexts may influence both participant behavior and the quality of the generated requirements artifacts. Third, the collaborative conditions were influenced by participant interactions and group dynamics, which may have affected both the elicitation process and the resulting requirements artifacts. Fourth, differences among conditions may reflect not only the elicitation approach but also the mechanisms used to produce the final requirements artifacts, particularly in the AI-assisted conditions. Fifth, the outputs produced by the Writer applet and the LLM-based conditions depend on prompt formulation and the behavior of the underlying language models. Consequently, variations in prompts or model responses may influence the generated artifacts. Finally, Strateegia primarily supports textual interactions. The absence of visual artifacts such as prototypes or diagrams may have influenced the elicitation process and the resulting requirements documents.

\section{Results}

\subsection{Requirements Artifact Quality}

The four requirements artifacts generated during the experiment were evaluated by two requirements engineering specialists using the quality criteria described in Section~\ref{sec:method}. Table~\ref{tab:artifact_quality} presents the mean scores assigned to each artifact across the evaluated dimensions.

\begin{table}[!ht]
\caption{Requirements artifact quality evaluation}
\label{tab:artifact_quality}
\centering
\footnotesize
\begin{tabular}{lcccc}
\toprule
Criterion & A & B & C & D \\
\midrule
Completeness & 3.5 & 4.5 & 4.0 & 5.0 \\
Clarity & 2.0 & 4.0 & 3.5 & 4.0 \\
Consistency & 3.0 & 4.0 & 3.5 & 4.0 \\
Absence of ambiguity & 2.5 & 3.5 & 3.5 & 3.5 \\
Level of detail & 2.5 & 3.0 & 3.0 & 3.5 \\
Traceability & 2.0 & 3.5 & 2.5 & 3.5 \\
Overall quality & 2.5 & 4.0 & 3.5 & 4.0 \\
\bottomrule
\end{tabular}
\end{table}

Condition A, corresponding to the collaborative process without AI support, received the lowest ratings across all evaluated quality dimensions. The artifact obtained scores of 3.5 for completeness, 2.0 for clarity, 3.0 for consistency, 2.5 for absence of ambiguity, 2.5 for detail, 2.0 for traceability, and 2.5 for overall quality. These results suggest that although the participants were able to identify relevant requirements through collaborative discussion, transforming these discussions into a structured and coherent requirements document remained challenging.

The three AI-supported approaches consistently received higher evaluations than the manually produced document. Condition B, which combined collaborative elicitation through Strateegia with AI-supported synthesis, received scores of 4.5 for completeness, 4.0 for clarity, 4.0 for consistency, 3.5 for absence of ambiguity, 3.0 for detail, 3.5 for traceability, and 4.0 for overall quality. Compared with Condition A, the largest improvements were observed in clarity and traceability. These differences suggest that AI-assisted synthesis may support the organization and consolidation of stakeholder contributions into more structured requirements artifacts.

Condition C, in which an LLM generated requirements directly from the problem description without stakeholder interaction, also produced comparatively positive results. The generated artifact received an overall quality score of 3.5 and obtained ratings above those of Condition A in all evaluated dimensions. This finding suggests that modern LLMs may be capable of generating reasonably complete and coherent initial requirements artifacts even when stakeholder discussions are not available. However, the scores remained below those obtained by the approaches that incorporated collaborative interaction.

The highest-rated artifact was produced under Condition D, where the transcript of a collaborative discussion was provided to the LLM. This artifact achieved the highest score for completeness (5.0) and detail (3.5), while also obtaining the highest or joint-highest scores for clarity, consistency, traceability, and overall quality. These results suggest that AI systems may benefit from access to information generated through stakeholder interaction, allowing the resulting artifact to reflect a broader set of requirements, concerns, and perspectives than those available through direct prompting alone.

A comparison between Conditions B and D reveals a particularly interesting pattern. Although the two approaches used different interaction mechanisms, both relied on collaborative stakeholder discussions followed by AI-supported synthesis and both received the highest overall evaluations. This pattern suggests that the combination of human collaboration and AI-based consolidation may be more important than the specific collaborative environment used to collect stakeholder contributions. In other words, the availability of rich discussion data may be a key factor supporting the generation of higher-quality requirements artifacts.

\subsection{Participant Perceptions of the Elicitation Process}

Participants involved in the collaborative conditions (A and B) completed a post-activity questionnaire assessing their perceptions of the elicitation process. Table~\ref{tab:process_quality} presents the mean responses for each questionnaire item.

\begin{table}[!ht]
\caption{Participant perceptions of the elicitation process}
\label{tab:process_quality}
\centering
\footnotesize
\begin{tabular}{p{6.0cm}cc}
\toprule
\textbf{Question} & \textbf{A} & \textbf{B} \\
\midrule
The process was clear and easy to understand & 3.3 & 5.0 \\
The process was easy to follow and execute & 4.3 & 5.0 \\
The process supported collaboration and coordination & 4.7 & 5.0 \\
The process required high mental effort & 3.0 & 3.3 \\
The tool was adequate and supported the activity & 5.0 & 4.7 \\
\bottomrule
\end{tabular}
\end{table}

Participants in the AI-supported collaborative condition reported consistently more positive perceptions regarding process clarity, ease of execution, and collaboration support. Process clarity increased from 3.3 in Condition A to 5.0 in Condition B, while ease of execution increased from 4.3 to 5.0. These results suggest that the structured workflow provided by Strateegia may have helped participants understand the sequence of elicitation activities and the objectives associated with each stage of the process.

Perceptions regarding collaboration were positive in both conditions, with mean scores of 4.7 and 5.0 for Conditions A and B, respectively. This result indicates that participants considered both approaches effective for supporting interaction and coordination among stakeholders. The slightly higher score observed in Condition B suggests that the platform structure may have contributed to organizing participant contributions and maintaining engagement throughout the elicitation process.

Reported cognitive effort was relatively similar across conditions. Participants in Condition A reported a mean score of 3.0, while participants in Condition B reported a mean score of 3.3. These values do not indicate a substantial difference between the approaches. Consequently, although participants perceived the AI-supported process as clearer and easier to follow, the results do not suggest a meaningful reduction in the mental effort associated with requirements elicitation.

Participants evaluated both approaches positively regarding tool adequacy. Condition A received a score of 5.0 and Condition B received 4.7. These results indicate that participants considered both environments appropriate for supporting the assigned task, although the introduction of AI support did not necessarily increase perceptions regarding the suitability of the tool itself.

\subsection{How does an AI-supported collaborative platform influence software requirements elicitation and the production of requirements artifacts compared with alternative elicitation approaches?}

AI-supported collaborative approaches produced the highest-rated requirements artifacts and were perceived as clearer and easier to execute than the traditional collaborative approach. Requirements artifacts generated from stakeholder discussions followed by AI-supported synthesis received higher evaluations for completeness, clarity, consistency, traceability, and overall quality than the manually produced artifact. The highest-rated artifacts were produced when AI was used to consolidate information generated through stakeholder collaboration, both in the Strateegia-supported condition and in the transcript-based condition. For practitioners, these results suggest that the primary contribution of AI in requirements elicitation is supporting the synthesis and documentation of stakeholder discussions rather than replacing stakeholder participation. Although direct prompting of an LLM produced a positively evaluated requirements artifact, the strongest results were obtained when AI had access to information generated through collaborative discussion. These findings suggest that combining stakeholder collaboration with AI-supported synthesis may support the production of more complete, clearer, and more consistent requirements documentation.
\section{Discussion}
\label{sec:discussion}

Recent studies have suggested that large language models can support requirements-related activities such as information extraction, interpretation, formalization, and artifact generation~\cite{arora2024advancing, abbasi2025towards}. Our findings are consistent with this literature, as all AI-supported approaches produced requirements artifacts that received higher evaluations than the manually produced document. However, the results also suggest that the quality of generated artifacts depends on the role assigned to AI within the process. While direct prompting produced a positively evaluated artifact, the highest-rated artifacts were obtained when AI synthesized information generated through stakeholder collaboration.

This observation reinforces the continued importance of stakeholder participation during elicitation activities. Traditional requirements elicitation approaches rely on interactions among stakeholders to capture needs, negotiate priorities, clarify assumptions, and develop a shared understanding of the problem space~\cite{nuseibeh2000requirements, sommerville2011software, pacheco2018requirements}. The highest-rated artifacts were generated when AI operated on information produced through stakeholder discussions, indicating that human interaction contributed contextual information, rationale, and perspectives that were not fully captured by the problem description alone.

The findings also provide a different perspective from that offered by many existing requirements engineering tools. Solutions such as DOORS Next, Jama Connect, ReqView, and Valispace primarily focus on requirements management, traceability, lifecycle support, and artifact maintenance, while aqua Cloud emphasizes AI-assisted writing and refinement. Comparatively less attention has been given to supporting the transition between collaborative stakeholder discussions and formal requirements documentation. Our findings suggest that this transition represents a useful opportunity for AI support.

The comparison between conditions further illustrates the complementary roles of humans and AI during requirements elicitation. Stakeholders contributed domain knowledge, contextual information, priorities, and negotiation outcomes, while AI contributed synthesis, organization, and documentation support. Neither the human-only condition nor the AI-only condition produced the highest-rated artifact. Instead, the strongest results emerged when collaboratively generated information was subsequently consolidated through AI.

The process-related findings reinforce this interpretation. Participants perceived the AI-supported collaborative approach as clearer and easier to execute than the traditional collaborative process, while collaboration remained positively evaluated in both conditions. These observations suggest that structured collaborative environments combined with AI-supported synthesis may help teams manage stakeholder contributions more systematically and transform discussion outcomes into formal requirements artifacts more effectively. Rather than replacing stakeholder participation, AI acted as a mechanism for consolidating discussions, organizing information, and producing structured requirements artifacts.

\section{Conclusions and Future Work}
\label{sec:conclusions}

This study explored the use of a collaborative platform supported by generative AI for software requirements elicitation. Our findings suggest that the combination of stakeholder collaboration and AI-supported synthesis represents a promising approach for producing requirements artifacts and supporting elicitation activities. The results indicate that AI can play a useful role in consolidating stakeholder contributions and transforming collaborative discussions into structured requirements documentation. This work contributes empirical evidence to ongoing discussions regarding the use of generative AI in software development and, more specifically, the role of human-AI collaboration during elicitation activities. Rather than functioning as a replacement for stakeholders, AI acted as a complementary mechanism for organizing, synthesizing, and documenting collaboratively generated knowledge. \textbf{Future Work.} Future studies should investigate the proposed approach using larger participant populations, additional evaluators, and multiple elicitation scenarios. Additional research could explore different language models, prompting strategies, and collaborative configurations, as well as extending AI support to activities such as requirements analysis, validation, prioritization, and management. Studies conducted in industrial environments would also provide additional insight into the practical adoption of AI-supported collaborative approaches and their integration into existing software development processes.

\ifCLASSOPTIONcaptionsoff
\newpage
\fi

\balance
\bibliographystyle{IEEEtran}
\bibliography{references}

\end{document}